\documentclass[pre,aps,byrevtex,preprint]{revtex4}
\usepackage{times}
\usepackage{epsf}
\usepackage{graphicx}
\usepackage{dcolumn}
\usepackage{bm}

\begin{document}

\title{Magnetophoresis of nonmagnetic particles in ferrofluids}
\author{Y. Gao, Y. C. Jian, L. F. Zhang, and J. P. Huang\footnote{Corresponding author.
E-mail: jphuang@fudan.edu.cn}} \affiliation{Surface Physics
Laboratory and Department of Physics, Fudan University, Shanghai
200433, China}

\begin{abstract}

Ferrofluids containing nonmagnetic particles are called inverse
ferrofluids. Based on the Ewald-Kornfeld formulation and the
Maxwell-Garnett theory, we theoretically investigate the
magnetophoretic force exerting on the nonmagnetic particles in
inverse ferrofluids due to the presence of a nonuniform magnetic
field, by taking into account the structural transition and
long-range interaction. We numerically demonstrate that the force
can be adjusted by choosing appropriate lattices, volume fractions,
geometric shapes, and conductivities of the nonmagnetic particles,
as well as frequencies of external magnetic fields.

\end{abstract}

\maketitle

%\date {\today}
%\address{Surface Physics Laboratory (National Key Laboratory)
%and Department of Physics, Fudan University, Shanghai 200433, China}
%\footnote{Email address: jphuang@fudan.edu.cn}
%\pacs{61.20.Gy, 75.50.Mm, 82.70 }

%%%%%%%%%%%%%%%%%%%%%%%%%%%%%%%%%%%%%%%%%%%%%%%%%

\section{Introduction}\label{intro}

Magnetizable particles experience a force in a nonuniform magnetic
field, which comes from the interaction between the induced magnetic
moment inside the particles and the external magnetic field. This
phenomenon is called magnetophoresis,$^{1,2}$ and has been exploited
in a variety of industrial and commercial process for separation and
beneficiation of solids suspended in liquids.$^3$ Magnetophoresis
may also open up a way to characterize and separate cells for
biochemical analysis based on intrinsic and extrinsic magnetic
properties of biological macromolecules.$^4$

Ferrofluids, also known as magnetic fluids, are a colloidal
suspension of single-domain ferromagnetic nanoparticles (e.g., made
of Fe$_3$O$_4$), with typical dimension around 10\,nm, dispersed in
a carrier liquid like water or kerosene.$^{3,5-7}$ Ferrofluids
containing nonmagnetic particles (or magnetic holes) are called
inverse ferrofluids. So far, inverse ferrofluids  have drawn much
attention for their potential industrial and biomedical
applications.$^{8-11}$ The size of the nonmagnetic particles (such
as polystyrene particles) is about 1$\sim$100 $\mu$m. Since the
nonmagnetic particles are much larger than the ferromagnetic
nanoparticles in host ferrofluids, the host can theoretically be
treated as a one-component continuum in which the much larger
nonmagnetic particles are embedded.$^{12,13}$ It is known that the
ground state of electrorheological fluids and magnetorheological
fluids is body-centered tetragonal (bct) lattices.$^{14,15}$ Due to
mathematical analogy, the ground state of inverse ferrofluids might
also be body-centered tetragonal (bct) lattices. Similar to
electrorheological fluids,$^{16}$ under appropriate conditions, such
as by adjusting the application of external magnetic fields,
structural transition may also appear in inverse ferrofluids, e.g.,
from the bct to the body-centered cubic (bcc), and then to the
face-centered cubic (fcc) lattices.

When an external magnetic field is applied, the nonmagnetic
particles  in inverse  ferrofluids can be seen to possess an
effective magnetic moment, which, however, is anti-parallel to the
magnetization of the host ferrofluid. In this sense, the nonmagnetic
particles also experience an equivalent magnetophoretic  force in
the presence of a nonuniform magnetic field naturally. In this work,
we shall apply a nonuniform magnetic field to investigate the
magnetophoresis of the nonmagnetic particles  in inverse ferrofluids
with different structures of specific lattices.

One of us$^{17}$ has investigated the dielectrophoretic force acting
on a microparticle in electrorheological solids under the
application of a nonuniform external electric field by using a
self-consistent method$^{18}$ and the spectral representation
approach. In our present work, we shall use the Ewald-Kornfeld
formulation and the Maxwell-Garnett theory for uniaxially
anisotropic suspensions to calculate the effective permeability of
the inverse ferrofluid, taking into account the long range
interaction arising from specific lattices. Based on this, the
magnetophoretic force acting on the nonmagnetic particles can be
calculated accordingly. We shall focus on the role of  types of
lattices, volume fractions, geometric shapes, and conductivities of
the nonmagnetic particles, as well as frequencies of external
fields.

This paper is organized as follows. In Section~\ref{Formalism}, we
first briefly review the theory of magnetophoresis
(Section~\ref{II-1}). By using the Ewald-Kornfeld formulation and
the Maxwell-Garnett theory for uniaxially anisotropic suspensions,
we calculate the effective permeability of the inverse ferrofluids
(Section~\ref{II-2}). Then we investigate magnetophoresis with eddy
current induction (Section~\ref{II-3}). In Section~\ref{results}, we
give the numerical results under different conditions. This paper
ends with a discussion and conclusions in Section~\ref{conclusion}.

\section{Formalism}\label{Formalism}

\subsection{Theory of magnetophoresis}\label{II-1}

To establish our theory, we first briefly review the theory of
magnetophoresis (for details of magnetophoresis, we refer the reader
to ref 1). Let us start by considering a homogeneous sphere with
radius $R$ and  magnetization $\mathbf{M}_{1}$ that is suspended in
a magnetically linear fluid of permeability $\mu_{2}$ and subjected
to a uniform magnetic field $\mathbf{H}_{0}$. The relation between
magnetic flux density $\mathbf{B}$,  magnetization $\mathbf{M}$, and
magnetic intensity $\mathbf{H}$ is given by
\begin{equation}
\mathbf{B}=\mu_{0}(\mathbf{H}+\mathbf{M}), \label{BHM}
\end{equation}
where $\mu_{0}$ is the permeability of free space. Eq~\ref{BHM} is
completely general, the  magnetization $\mathbf{M}$ includes
permanent magnetization plus any linear or nonlinear function of
$\mathbf{H}$. $\mathbf{M}_1$ is assumed to be parallel to
$\mathbf{H}_{0},$ $\mathbf{M}_{1}\|\mathbf{H}_{0}$. We also assume
that no electric current flows anywhere. Thus, there is
\begin{equation}
\nabla\times\mathbf{H}=0
\end{equation}
everywhere. Then, this magnetostatics problem may be solved using
a scalar potential $\psi$, defined by $\mathbf{H}=-\nabla\psi$.
The general solutions for $\psi_{1}$ and $\psi_{2}$ outside and
inside the sphere are given by respectively,
\begin{eqnarray}
\psi_{1}(r,\theta) &=& -H_{0}r \cos\theta+\frac{a
\cos\theta}{r^2},\,\,\,
r>R, \label{out}\\
 \psi_{2}(r,\theta) &=& -br\cos\theta,\,\,\, r<R.
\end{eqnarray}
The coefficients $a$ and $b$ can be determined by considering the
appropriate boundary conditions,
\begin{eqnarray}
a &=& \frac{\mu_{0}-\mu_{2}}{\mu_{0}+2
\mu_{2}}R^3H_{0}+\frac{\mu_{0}}{\mu_{0}+2 \mu_{2}}R^3M_{1},\\
b &=& \frac{3\mu_{2}}{\mu_{0}+2
\mu_{2}}H_{0}-\frac{\mu_{0}}{\mu_{0}+2\mu_{2}}M_{1}.
\end{eqnarray}
It is worth noting that $b$ denotes the magnitude of the uniform
magnetic field $H_{1}$ inside the sphere. We can determine the
effective magnetic dipole moment $\mathbf{m}_{\mathrm{eff}}$ by
comparing the dipole term in  eq~\ref{out} with
\begin{equation}
\psi_{\mathrm{dipole}}=\frac{m_{\mathrm{eff}}\cos\theta}{4 \pi
r^2}.
\end{equation}
Then, the effective magnetic moment is given by
\begin{equation}
\mathbf{m}_{\mathrm{eff}} = 4 \pi R^3
\left[\frac{\mu_{0}-\mu_{2}}{\mu_{0}+2
\mu_{2}}\mathbf{H}_{0}+\frac{\mu_{0}}{\mu_{0}+2
\mu_{2}}\mathbf{M}_{1}\right].
\end{equation}
For a magnetically linear particle of radius $R$ and permeability
$\mu_{1}$ suspended in a host medium with permeability $\mu_{2}$,
we have
\begin{equation}
\mathbf{M}_{1}=\chi_{1}\mathbf{H}_{1},
\end{equation}
where $\chi_{1}=\mu_{1}/\mu_{0}-1$ is the susceptibility of the
particle. In this case, the effective magnetic dipole moment
vector $\mathbf{m}_{\mathrm{eff}}$ takes a form very similar to
that for the effective dipole moment of dielectric spheres,
\begin{equation}
\mathbf{m}_{\mathrm{eff}}=4\pi R^{3}K(\mu_{1},\mu_{2})\mathbf{H}_{0},
\label{L-MAP}
\end{equation}
where the Clausius-Mossotti factor $K(\mu_{1},\mu_{2})$ is given by,
\begin{equation}
K(\mu_{1},\mu_{2})=\frac{\mu_{1}-\mu_{2}}{\mu_{1}+2\mu_{2}}.
\end{equation}
The magnetophoretic force exerted by a nonuniform magnetic field
on a magnetic dipole can then be written as
\begin{equation}
F_{\mathrm{dipole}}=\mu_{2}\mathbf{m}_{\mathrm{eff}}\cdot{\bf
\nabla} H_{0}.
\end{equation}
Combining this expression with that for the effective magnetic
dipole moment eq~\ref{L-MAP}, the magnetophoretic force exerted on a
magnetizable spherical particle in a nonuniform magnetic field is
written as$^{1}$
\begin{equation}
F_{\mathrm{MAP}}=2\pi\mu_{2}R^{3}K(\mu_{1},\mu_{2})\nabla
H_{0}^{2}. \label{L-MAP-F}
\end{equation}
Eq~\ref{L-MAP-F} is similar to the expression of the
dielectrophoretic force on dielectric particles.$^{1,19}$ It is
apparent to see that the magnetophoretic force are proportional to
the particle volume and the permeability of the suspension medium.
And the force is directed along the gradient of the external
magnetic field. There are two kinds of magnetophoresis depending
upon the relative magnitudes of $\mu_{1}$ and $\mu_{2}$. If $K>0$
(namely, $\mu_{1}>\mu_{2}$), particles are attracted to magnetic
field intensity maxima and repelled from minima. This phenomenon
is called positive magnetophoresis. On the contrary, if $K<0$
(i.e., $\mu_{1}<\mu_{2}$), inverse behavior appears, which is
called negative magnetophoresis. Generally, there may be two
magnetophoresis systems. One is that ferromagnetic particles are
suspended in a medium that can be either nonmagnetic or less
(linear) magnetizable, and the other one is that nonmagnetic
particles are in a magnetizable liquid like ferrofluids. As for
the second case of our interest in this work, in case of a
moderate magnetic field (satisfying the assumption of a
magnetically linear host), we can use eq~\ref{L-MAP-F} by setting
$\mu_{1}=\mu_{0}$ to calculate the magnetophoretic force on the
nonmagnetic particles submerged in a host ferrofluid.

\subsection{Effective permeability}\label{II-2}

We resort to the Ewald-Kornfeld formulation to calculate the
effective permeability by including the structural transition and
corresponding long-range interaction  explicitly. A bct lattice
can be regarded as a tetragonal lattice, plus a basis of two
particles each of which is fixed with an induced point magnetic
dipole at its center. One of the two particles is located at a
corner and the other one at the body center of the tetragonal unit
cell. The tetragonal lattice has a lattice constant $c_{1}$= $\xi
q$ along the $z$ axis and lattice constants $a_{1}=b_{1}=\xi
q^{-\frac{1 }{2}}$ along $x$ and $y$ axes, respectively. In this
paper, we identify a different structure transformation from bct
to bcc, and to fcc lattices by changing the uniaxial lattice
constant $c_{1}$ under the hard-sphere constraint. The volume of
the tetragonal unit cell $V_c$ remains unchanged (i.\,e., $V_c =
\xi^{3}$) as $q$ varies. In this way, the degree of anisotropy of
the tetragonal lattices is measured by how $q$ is deviated from
unity and the uniaxial anisotropic axis is along the $z$ axis. In
particular, $q=0.87358$, $1,$ and $2^{\frac{1}{3}}$ represent the
bct, bcc, and fcc lattices, respectively.

When an external magnetic field $\mathbf{H}_{0}$ is directed along
the $x$ axis, the induced magnetic moment $\mathbf{m}_{\rm eff}$ is
perpendicular to the uniaxial anisotropic  axis ($z$ axis). At the
lattice point $\mathbf{R}={\bf 0}$, the local field $
\mathbf{H}_{L}$ can be determined by using the Ewald-Kornfeld
formulation,$^{20-22}$
\begin{eqnarray}
\mathbf{H}_{L}&=&\mathbf{m}_{\rm
eff}\sum_{j=1}^{2}\sum_{\mathbf{R}\neq 0
}[-\gamma_{1}(R_{j})+x_{j}^{2}q^{2}\gamma_{2}(R_{j})]  \nonumber \\
&&-\frac{4\pi\mathbf{m}_{\rm eff}}{v_{c}}\sum_{\mathbf{G}\neq 0}
\Pi(\mathbf{G})\frac{
G_{x}^{2}}{G^{2}}\exp\left(\frac{-G^{2}}{4\eta^{2}}\right)
+\frac{4\mathbf{m}_{\rm eff} \eta^{3}}{3\sqrt{\pi}}.  \label{EKF}
\end{eqnarray}
where $\gamma_{1}$ and $\gamma_{2}$ are two coefficients, given by
\begin{eqnarray}
\gamma_{1}(r)&=&\frac{erfc(\eta r)}{r^{3}}+\frac{2\eta}{\sqrt{\pi}r^{2}}%
\exp(-\eta^{2}r^{2}),\nonumber\\
\gamma_{2}(r)&=&\frac{3erfc( \eta
r)}{r^{5}}+(\frac{4\eta^{3}}{\sqrt{\pi}r^{2}}
+\frac{6\eta}{\sqrt{\pi}r^{4}})\exp-(\eta^{2}r^{2}),\nonumber
\end{eqnarray}
where erfc$(\eta r)$ is the complementary error function, and $\eta$
is an adjustable parameter making the summations in real and
reciprocal lattices converge rapidly. In eq~\ref{EKF}, $\mathbf{R}$
and $\mathbf{G}$ denote the lattice vector and reciprocal lattice
vector, respectively,

where $l, m, n, u, v,$ and $w$ are integers. In addition, $x_{j}$
and $R_{j}$ in eq~\ref{EKF} are given by
\begin{eqnarray}
x_{j} &=& l-\frac{j-1}{2},\nonumber\\
R_{j} &=&|\mathbf{R}-\frac{j-1}{2}(a_{1}\hat{\mathbf{x}}
+a_{1}\hat{\mathbf{y}}+c_{1}\hat{\mathbf{z}})|,\nonumber
\end{eqnarray}
and $\Pi(\mathbf{G})=1+\exp[i(u+v+w)/\pi]$ is the structure
factor. The local magnetic field can be obtained by summing over
all integer indices, $(u, v, w) \neq (0, 0, 0)$ for the summation
in the reciprocal lattice, and $(j, l, m, n) \neq (1, 0, 0, 0)$
for the summation in the real lattice. In our calculation, because
of the exponential factors, we may impose an upper limit to the
indices, i.e., all indices ranging from $-L$ to $L$, where $L$ is
a positive integer. We consider an infinite lattice. So, for
finite lattices, one must be careful about the effects of
different boundary conditions.

Now, let us define a local field factor $\alpha$ as
\begin{equation}
\alpha=\frac{3}{4\pi}\frac{\mathbf{H}_{L}}{\sum_{V_c}\mathbf{m}_{\rm
eff}/V_c}=\frac{3V_{c}}{8\pi}\frac{\mathbf{H}_{L}}{\mathbf{m}_{\rm
eff}},
\end{equation}
where  $\sum_{V_c}\mathbf{m}_{\rm eff}=2{\bf m}_{\rm eff}$ means the
sum of the two vector magnetic moments  in the unit cell. It is
worth remaking that $\alpha$ is a function of single variable $q$,
and that it stands for $\alpha_{T}$ (local field factor in
transverse field cases) and $\alpha_{L}$ (local field factor in
longitudinal field cases) which satisfy a famous sum rule in
condensed matter physics for $\alpha_{T}$ and $\alpha_{L}$,
$2\alpha_{T}+\alpha_{L}=3$.$^{18,19}$ Here the longitudinal
(transverse) field case corresponds to the fact that the external
magnetic field is parallel (perpendicular) to the uniaxial
anisotropic $z$ axis. Next, for calculating the effective
permeability, we  use the Maxwell-Garnett theory for uniaxially
anisotropic suspensions,$^{23,24}$
\begin{equation}
\frac{\mu_{e}-\mu_{2}}{\alpha_{L,T}\mu_{e}+(3-\alpha_{L,T})\mu_{2}}=f\frac{%
\mu_{1}-\mu_{2}}{\mu_{1}+2\mu_{2}},  \label{MGE}
\end{equation}
where $\mu_{e}$ stands for the effective permeability of the
inverse ferrofluid, $\mu_{1}$  the permeability of the nonmagnetic
particles, namely $\mu_1 = \mu_0,$ $\mu_{2}$  the permeability of
the host ferrofluid, and $f$ the volume fraction of the
nonmagnetic particles  in the inverse ferrofluids.

It should be noted that eq~\ref{MGE} holds for the case of
nonmagnetic particles with a spherical shape only. For nonspherical
cases (e.g., prolate or oblate spheroids), we may use
 demagnetizing factor $g$ to describe the shape of different kinds of
spheroids. For a prolate spheroid with three principal axes, $a$,
$b$, and $c,$ there is $a=b<c,$ and the
 demagnetizing factor ${g}_{L}$ along the major axis
$c$ is$^{25,26}$
\begin{equation}
{g}_{L}=\frac{1}{k^{2}-1}\left[\frac{k}{\sqrt{k^{2}-1}}\ln(k+\sqrt{k^{2}-1}%
)-1\right],
\end{equation}
where $k=c/a>1$ and ${g}_{L}<1/3$. For an oblate spheroid with
$a<b=c$, the  demagnetizing factor ${g}_{L}$ along the major axis
$a$ is$^{25,26}$
\begin{equation}
{g}_{L}=1-\left[\frac{k^{2}}{(k^{2}-1)^\frac{3}{2}}\arcsin\frac{\sqrt{k^{2}-1
}}{k}-\frac{1}{k^{2}-1}\right],
\end{equation}
where $k=c/a>1$ and ${g}_{L} > 1/3.$ For both cases, there is a
geometrical sum rule for ${g}_{L}$ and ${g}_{T}$,
$2{g}_{T}+{g}_{L}=1$, where ${g}_{T}$ denotes the demagnetizing
factor along the minor axis. Note that the spherical shape of
nonmagnetic particles corresponds to ${g}_{L}={g}_{T}=1/3.$ So far,
to include the nonspherical shape effects, erq~\ref{MGE} for
calculating the effective permeability of inverse ferrofluids
containing spherical nonmagnetic particles might be modified as
\begin{equation}
\frac{\mu_{e}-\mu_{2}}{\mu_{2}+\frac{\alpha_{L,T}}{3}(\mu_{e}-\mu_{2})}=
f\frac{\mu_{1}-\mu_{2}}{\mu_{2}+g_{L,T}(\mu_{1}-\mu_{2})}.
\label{MGA}
\end{equation}
Apparently, the substitution of ${g}_{L}={g}_{T}=1/3$ into
eq~\ref{MGA} reduces to eq~\ref{MGE}.

\subsection{Magnetophoresis with eddy current induction}
\label{II-3}

When an ac (alternating current)  magnetic field is applied to an
electrically conductive magnetizable particle, eddy currents are
induced, which tend to oppose penetration of the field into the
particle.  At sufficiently high frequencies, these eddy currents
strongly influence the effective magnetic moment. Within the
particle, $\nabla \times \mathbf{H}\neq 0.$ Similar derivation of
this section can be found in ref~1 for a magnetic particle in a
linear medium. Nevertheless, here we discuss a different system,
thus yielding the following formulae with some difference.

Let us consider a nonmagnetic conductive particle  with an effective
magnetic dipole of magnitude $m_{\rm eff}$ [note that in this case
there is ${\bf m}_{\rm eff}\|(-{\bf H}_0)$ as mentioned in
Section~\ref{intro}], which is embedded in the system containing
both a magnetically linear host ferrofluid and many other
nonmagnetic conductive particles. In this case, the nonmagnetic
particles are also assumed to situate in specific lattices and exist
in the shape of spheres. The effective permeability of the whole
system $\mu _{e}$ can be calculated according to eq~\ref{MGE}. The
magnetic vector potential ${\bf \Psi}$  at position $\mathbf{r}$ may
be written as
\begin{equation}
{\bf \Psi}=\frac{\mu _{e}\mathbf{m}_{\mathrm{eff}}\times
\mathbf{r}}{4\pi r^{3}},  \label{VPA}
\end{equation}
where ${\bf \nabla} \times {\bf \Psi}=\mathbf{B}$, $\mathbf{B}$ is
the magnetic flux density. If we align the moment anti-parallel with
$z$ axis, that is $
\mathbf{m}_{\mathrm{eff}}=-m_{\mathrm{eff}}\hat{\mathbf{z}}.$ Then
eq~\ref{VPA} can be rewritten as
\begin{equation}
{\bf \Psi}=-\frac{\mu _{e}m_{\mathrm{eff}}\sin \theta }{4\pi
r^{2}}\hat{{\bf \phi}},
\end{equation}
where $\hat{{\bf \phi}}$ is the azimuthal unit vector in spherical
coordinates. We consider a conductive nonmagnetic spherical
particle of permeability $\mu _{1}(=\mu_0)$ and conductivity
$\sigma _{1}$, which is subjected to a uniform, linearly polarized
magnetic field $\mathbf{H} (t)=H_{0}\cos \omega
t\hat{\mathbf{z}}.$ The effective moment of the sphere can be
written as
\begin{equation}
\mathbf{m}_{\mathrm{eff}}=-4\pi R^{3}L(\omega
)H_{0}\hat{\mathbf{z}},
\end{equation}
where $L(\omega )=R^{3}D/2$ is a frequency-dependent magnetization
coefficient, and $D$ depends on $\mu _{e}$, $\mu _{1}$, $\sigma
_{1}$, and $\omega $,
\begin{equation}
D=\frac{(2\mu _{1}+\mu _{e})\nu I_{-1/2}-[\mu _{e}(1+\nu
^{2})+2\mu _{1}]I_{+1/2}}{(\mu _{1}-\mu _{e})\nu I_{-1/2}+[\mu
_{e}(1+\nu ^{2})-\mu _{1}]I_{+1/2}},
\end{equation}
with $I_{\pm 1/2}=I_{\pm (\nu )}$ and $\nu =\sqrt{j\mu _{1}\sigma
_{1}\omega R}$. The quantities $I_{\pm 1/2}$ are half-integer
order, modified Bessel functions of the first kind. For a
nonuniform ac magnetic field, according to the effective dipole
method (see Sections~\ref{II-1}~and~\ref{II-2}), the time-average
magnetophoretic force acting on a nonmagnetic conductive particle
is
\begin{equation}
\langle F_{\mathrm{MAP}}(t)\rangle =\frac{\mu _{e}}{2}{\rm
Re}[\mathbf{m}_{\mathrm{eff}}\cdot {\bf \nabla}H].
\end{equation}
Next, we can obtain the expression for the magnetophoretic force
in terms of the real part of $L$,
\begin{equation}
\langle F_{\mathrm{MAP}}(t)\rangle =-2\pi \mu _{e}R^{3}{\rm
Re}[L]\nabla H_{0}^{2}.
\end{equation}
A simplified expression for ${\rm Re}[L]$ may be obtained by
invoking certain Bessel function identities,
\begin{equation}
{\rm Re}[L]=\frac{3\mu _{R}}{2}\frac{(\mu
_{R}-1)B^{2}+R_{m}^{2}(\sinh 2R_{m}-\sin 2R_{m})A}{(\mu
_{R}-1)^{2}B^{2}+R_{m}^{2}A^{2}}-\frac{1}{2},
\end{equation}
where $\mu _{R}=\mu _{1}/\mu _{e}$ is the relative permeability,
$R_{m}=R\sqrt{\omega \mu _{1}\sigma _{1}/2}$ a dimensionless
modulus called magnetic Reynolds number, and $A$ and $B$ are
respectively given by
\begin{eqnarray}
A&=&2R_{m}(\cosh 2R_{m}-\cos 2R_{m})+(\mu _{R}-1)(\sinh
2R_{m}-\sin2R_{m}),\nonumber\\
B&=&\cosh 2R_{m}-\cos 2R_{m}-R_{m}(\sinh 2R_{m}+\sin
2R_{m}).\nonumber
\end{eqnarray}

\section{Results}\label{results}

If the shape of the nonmagnetic particle is spherical, the
magnetophoretic force on the nonmagnetic particle
$F_{\mathrm{MAP}}$ can be calculated by
\begin{eqnarray}
F_{\mathrm{MAP}} &=& 2\pi\mu_{e}R^{3}K(\mu_{1},\mu_{e})\nabla
H_{0}^{2},\label{MAP-2}\\
K(\mu_{1},\mu_{e}) &=& \frac{\mu_{1}-\mu_{e}}{\mu_{1}+2\mu_{e}}.
\end{eqnarray}
For obtaining eq~\ref{MAP-2}, in order to include the lattice
effect, $\mu_{2}$ in eq~\ref{L-MAP-F} has been replaced by $\mu_{e}$
that can be calculated according to eq~\ref{MGE}. During the
numerical calculation, for convenience, $F$ stands for
$-F_{\rm{MAP}}/(2\pi R^{3}\nabla H_{0}^{2}).$  In our calculation,
we choose the parameters $\mu_{1}=\mu_0$ and $\mu_{2}=2.9\mu_0 .$ In
this case, the parameter $K(\mu_{1},\mu_{e})<0$  corresponds to the
phenomenon of negative magnetophoresis.

Figure~\ref{fig1} displays the dependency of $F$ on $q$ when the
external nonuniform magnetic field is parallel to the uniaxial
anisotropic axis (longitudinal field cases) for three different
volume fractions $f=0.15,$ $f=0.25,$ and $f=0.35$. While
Figure~\ref{fig2} shows the dependency of $F$ on $q$ when the
external nonuniform magnetic field is perpendicular to the uniaxial
anisotropic axis (transverse field cases) for the same three
different volume fractions.  Because of the existence of anisotropy,
Figure~\ref{fig1} displays that $F$ decreases as $q$ increases.
Inverse behavior appears in Figure~\ref{fig2}. It is apparent that
the lattice effect plays an important role in magnetophoresis of
nonmagnetic particles in inverse ferrofluids. In the mean time,
increasing volume fraction $f$ of nonmagnetic particles leads to
decreasing normalized magnetophoretic forces for both longitudinal
and transverse field cases. As the volume fraction of the
nonmagnetic particles increases, the lattice effect becomes
stronger. By using the  Maxwell-Garnett theory for uniaxially
anisotropic suspensions eq~\ref{MGE}, a smaller effective
permeability of the inverse ferrofluid can be obtained.
Consequently, the absolute value of the normalized magnetophoretic
force on the nonmagnetic particles becomes smaller.

The inclusion of the shape effects makes the magnetophoretic force
expression more complicated for a nonspherical particle than for a
spherical. For  prolate spheroidal particles ($a=b<c$), which are
aligned with its major axis being parallel [corresponding to the
case of  $\alpha_L$ in eq~\ref{MGA} and $g_L$ in eqs~\ref{MGA} and
\ref{N-MAP}] or perpendicular [reflected by $\alpha_T$ in
eq~\ref{MGA} and $g_T$ in eqs~\ref{MGA} and \ref{N-MAP}] to the
direction of the external nonuniform magnetic field, the
magnetophoretic force expression resembles that for a sphere,
\begin{equation}
F_{\mathrm{MAP}}=2 \pi a^{2}c \left[\frac{1}{3}\frac{(\mu_{1}-\mu_{e})}{1+(%
\frac{\mu_{1}-\mu_{e}}{\mu_{e}}){g}_{L,T}}\right]\nabla H_{0}^{2},
\label{N-MAP}
\end{equation}
where $\mu_{e}$ should be calculated according to eq~\ref{MGA} by
including the lattice effects. From this equation, we  find that the
force remains proportional to the scalar product of particle volume
and excess permeability. Ellipsoidal particles exhibit the same
positive and negative magnetophoresis phenomenon, depending on
whether $\mu_{1}$ is greater than or less than $\mu_{e}$. This
expression eq~\ref{N-MAP} is also valid for oblate spheroids
($a<b=c$) by replacing $a^2c$ with $ac^2$. In fact,  a particle with
its short axis (e.g., $a$ axis for the oblate spheroid discussed
herein) parallel to the magnetic field is in a very unstable
alignment. In this work, for completeness of including shape
effects, we shall also discuss such a case of oblate spheroidal
particles with its short axis (i.e., major axis or $a$ axis)
parallel to the magnetic field. For a spherical particle, there is
$a=b=c$ and $g_L=g_T=1/3$, and eq~\ref{N-MAP} reduces to
eq~\ref{MAP-2} naturally.

Next, for convenience, we denote $F_{1}$ as $-F_{\rm{MAP}}/({2\pi
a^{2}c}\nabla H_{0}^{2})$ for prolate spheroids or
$-F_{\rm{MAP}}/({2\pi ac^{2}}\nabla H_{0}^{2})$ for oblate
spheroids, which means that we only calculate the normalized
magnetophoretic force.

Figure~\ref{fig3} displays the dependency of $F_{1}$ on $q$ when the
external nonuniform magnetic field is parallel to the uniaxial
anisotropic axis (longitudinal field cases) for different aspect
ratios $k$ for prolate spheroidal cases. The normalized
magnetophoretic force of the nonmagnetic particle decrease as $q$
($q>0.7$) or $k$ increases. Nevertheless, at small $q$, $0.6<q<0.7$,
the situation is complicated, but the difference for different $k$
is very small. On the other hand, for the transverse field cases
shown in Figure~\ref{fig4}, as $q$ and/or $k$ increases, the
normalized magnetophoretic force is caused to increase for the full
range of $q$.

Figure~\ref{fig5} displays the dependency of $F_{1}$ on $q$ when the
external nonuniform magnetic field is perpendicular to the uniaxial
anisotropic axis (transverse field cases) for different aspect
ratios $k$ for oblate spheroidal cases. The normalized
magnetophoretic force of the nonmagnetic particle decreases as $q$
increases, and increases as the aspect ratio $k$ increases. Inverse
behavior appears for the transverse field cases, which is plotted in
Figure~\ref{fig6}. In addition, the relationship between $F_{1}$ and
the volume fraction of the nonmagnetic particles is also
investigated (no figures shown here). The normalized magnetophoretic
force on the nonmagnetic particle is found to decrease as the volume
fraction increases, which are consistent with the result of the
spherical case.

In section~\ref{II-3}, we consider the magnetophoresis with eddy
current induction. The conductivity of the nonmagnetic particle
(such as polystyrene) can be changed  by using the method of
doping. Again, for convenience,  $F_{2}$ stands for the normalized
magnetophoretic force, $F_2 = -\langle
F_{\mathrm{MAP}}(t)\rangle/2\pi R^3 \nabla H_{0}^{2}.$
Figure~\ref{fig7} displays the dependency of $F_{2}$ on different
frequencies $\omega$ for different conductivities $\sigma$. It is
shown that the normalized magnetophoretic force of the nonmagnetic
particle decrease as $\omega$ and/or $\sigma$ increase for the
current parameters in use.

\section{Discussion and conclusion}\label{conclusion}

We have theoretically investigated the magnetophoretic force on the
nonmagnetic particles  suspended in a host ferrofluid due to
nonuniform magnetic fields, by taking into account the effects of
structural transition and corresponding long-range interaction. We
have resorted to the Ewald-Kornfeld formulation and the
Maxwell-Garnett theory for calculating the effective permeability of
the inverse ferrofluids. And we have also investigated the role of
the volume fraction, geometric shape, and conductivity of the
nonmagnetic particles, as well as frequencies of external fields.

In this work, we have plotted the figures by using the values for
the volume fraction, $f=0.15, 0.25$ and $0.35,$ which actually
corresponds to low concentration systems. Nevertheless, it would be
realizable as the nonmagnetic particles can have a solid hard core
and a relative soft coating (e.g., by attaching long polymer chains,
etc.) to avoid aggregation. Thus the model of a soft particle with
an embedded point dipole has been used throughout this work. In this
case, the many-body (local-field) effects are of importance, while
the multipolar effects can be neglected. This is exactly part of the
emphasis in this work.  As the nonmagnetic particles are located
closely, the multipolar interaction between them is expected to play
a role.$^{27,28}$ In this regard, it is of value to extend the
present work to investigate the effect of  multipolar interaction.
In addition, it is also interesting to see what happens if the
particle is superconductor. In the Appendix, we develop the theory
for describing the magnetophoretic force acting on type I
superconducting particles. In all our figures, only the normalized
magnetophoretic forces are displayed because it is very complicated
to calculate the field gradient $\nabla H_{0}^{2}$ for the lattices
of our discussion accurately. In so doing, the contribution of the
gradient has not been reflected in the figures. Fortunately, the
difference of the gradient for the different lattices can be very
small due to the closely packing of the nonmagnetic particles in the
lattices.

The present theory is valid for magnetically linear media, e.g.,
ferrofluids in a moderate magnetic field. Throughout the paper,
for convenience  we have assumed the host ferrofluid to behave as
an isotropic magnetic  medium (continuum) even though a magnetic
field is applied. In fact,  an external magnetic field can also
induce a host ferrofluid to be anisotropic, and hence the degree
of anisotropy in the host depends on the strength of the external
field accordingly. Also, in Section~\ref{II-3}, we have assumed
that only the suspended nonmagnetic particles are conducting while
the host is not. Therefore, it is also interesting to investigate
the cases of an anisotropic and/or conducting host.

For magnetic materials, nonlinear responses can occur for a high
magnetic field. In case of moderate field, linear responses are
expected to dominate. The latter just corresponds to the case
discussed in this work. We have also investigated magnetophoresis
with eddy current induction. In fact, conductive particles can
exhibit diamagnetic responses in ac fields because of induced eddy
currents. Nevertheless, the magnetic susceptibilities of some
conductive (diamagnetic) particles like silver or copper are
generally $-10^{-5}$ or so. Thus, their relative magnetic
permeabilities could be $1 + (-10^{-5}) \approx 1$, as already
used for Fig.~\ref{fig7}.

For treating practical application, the present lattice
theoretical model should be replaced by the random model, mainly
due to the existence of a random distribution of particles. In so
doing, some effective medium approximations may be used instead,
e.g., the Maxwell-Garnett approximation, the Bruggeman
approximation, and so on.

To sum up, our results have shown that the magnetophoretic force
acting on a nonmagnetic particle suspended in a host ferrofluid
due to the existence of a nonuniform magnetic field can be
adjusted significantly by choosing appropriate  lattices, volume
fractions, geometric shapes and conductivities  of the nonmagnetic
particles, as well as frequencies of external fields.

%%%%%%%%%%%%%%%%%%%%%%

\section*{Acknowledgements}

We acknowledge the financial support by the National Natural Science
Foundation of China under Grant No.~10604014,  by Chinese National
Key Basic Research Special Fund under Grant No. 2006CB921706, by the
Shanghai Education Committee and the Shanghai Education Development
Foundation (Shuguang project), and by the Pujiang Talent Project
(No. 06PJ14006) of the Shanghai Science and Technology Committee.
Y.G. greatly appreciates Professor X. Sun of Fudan University for
his generous help. We thank Professor R. Tao of Temple University
for fruitful collaboration on the discovery that bct lattices are
the ground state of inverse ferrofluids.

% We thank Professor K. W. Yu of The Chinese University of Hong Kong for his fruitful discussion.

\clearpage
\newpage

\section*{Appendix: Magnetophoresis of type I superconducting particles}

A conventional type I superconducting particle exhibiting the
Meissner effect may be modelled as a perfect diamagnetic object.
Such a model is adequate as long as the magnitude of a magnetic
field  does not exceed the critical field strength of the material.
If these limits are not exceeded, we may estimate the magnetic
moment $\mathbf{m}_{\mathrm{eff}}$ for a type I superconducting
particle by taking the $K(\mu_1,\mu_2) =-0.5$ limit of
eqs~\ref{L-MAP}~and~\ref {L-MAP-F}, and then obtain the effective
dipole moment $\mathbf{m}_{\mathrm{eff}}$
\begin{equation}
\mathbf{m}_{\mathrm{eff}}=-2\pi R^3\mathbf{H}_{0},
\end{equation}
and the magnetophoretic force $F_{\mathrm{MAP}}$
\begin{equation}
F_{\mathrm{MAP}}=-\pi\mu_{e}R^3\nabla H_{0}^2.  \label{S-MAP}
\end{equation}
Eq~\ref{S-MAP} indicates that type I superconducting particles
will always seek minima of the external applied  magnetic field
and can be levitated passively in a cusped magnetic field. As a
matter of fact, eq~\ref{S-MAP} is developed from that (see ref~1)
for a single type I superconducting particle by taking into
account the lattices effects or, alternatively, introducing the
parameter of effective permeability $\mu_e$. It is known that the
diamagnetic behavior is very strong in a type I superconductor.
For such a superconducting particle, which is perfectly
diamagnetic, its magnetic susceptibility is $-1$, which causes its
relative magnetic permeability to be $0$. Nevertheless, this does
not affect the validity of the formulae presented herein.

\newpage
%\begin{references}
\section*{References and Notes}

(1) Jones, T. B. , \textit{Electromechanics of Particles}, Cambridge
University Press: New York, 1995; Chap.~III.

(2) Sahoo, Y.; Goodarzi, A.; Swihart, M. T. ; Ohulchanskyy, T. Y.;
Kaur, N.; Furlani, E. P.; Prasad, P. N. \textit{J. Phys. Chem. B}
{\bf 2005}, 109, 3879.

(3) Odenbach, S. {\it Magnetoviscous Effects in Ferrofluids};
Springer: Berlin, 2002.

(4) Zborowski, M.; Ostera, G. R.; Moore, L. R.; Milliron, S.;
Chalmers, J. J.; Schechter, A. N.; \textit{Biophys. J.}
\textbf{2003}, 84,2638.

(5) Rosenweig, R. E. \textit{Ferrohydrodynamics}; Cambridge
University Press: Cambridge, 1985.

(6) Huang, J. P. \textit{J. Phys. Chem. B } {\bf 2004}, 108, 13901.

(7) M\'{e}riguet, G.; Cousin, F.; Dubois, E.; Bou\'{e}, F.; Cebers,
A.; Farago, B.; Perzynski, R. \textit{J. Phys. Chem. B } {\bf 2006},
110, 4378.

(8) Skjeltorp, A. T.  \textit{Phys. Rev. Lett} \textbf{1983}, 51,
2306.

(9) Ugelstad, J.; Stenstad, P.; Kilaas, L.; Prestvik, W. S.; Herje,
R.; Berge, A.; Hornes, E. \textit{Blood Purif} \textbf{1993}, 11,
349.

(10) Hayter,J. B.; Pynn, R.; Charles, S.; Skjeltorp, A. T.;
Trewhella, J.; Stubbs, G.; Timmins,  P. \textit{Phys. Rev. Lett.}
\textbf{1989}, 62, 1667.

(11) Koenig, A.; H¡äebraud, P.; Gosse, C.; Dreyfus, R.; Baudry, J.;
Bertrand, E.; Bibette, J. \textit{Phys. Rev. Lett.} {\bf 2005}, 95,
128301.

(12) Feinstein,  E.; Prentiss, M. \textit{J. Appl. Phys.}
\textbf{2006}, 99, 064910.

(13) de Gans, B. J.; Blom, C.; Philipse, A. P.; Mellema, J.
\textit{Phys. Rev. E} \textbf{1999}, 60, 4518.

(14) Tao, R.; Sun, J. M.  \textit{Phys. Rev. Lett.} \textbf{1991},
67, 398.

(15) Zhou, L.; Wen, W. J.; Sheng, P. \textit{Phys. Rev. Lett.}
\textbf{1998}, 81, 1509.

(16) Lo, C. K.; Yu, K. W. \textit{Phys. Rev. E} \textbf{2001}, 64,
031501.

(17) Huang, J. P.  \textit{Chem. Phys. Lett.} \textbf{2004}, 390,
380.

(18) Bergman, D. J. \textit{Phys. Rep.} \textbf{1978}, 43, 377.

(19) Pohl, H. A. {\it Dielectrophoresis}; Cambridge University
Press: Cambridge, 1978.

(20) Huang, J. P. \textit{Phys. Rev. E} \textbf{2004}, 70, 041403.

(21) Ewald, P. P. \textit{Ann. Phys.} (Leipzig) \textbf{1921}, 64,
253; Kornfeld, H. \textit{Z. Phys.} \textbf{1924}, 22, 27.

(22) Huang, J. P.; Yu, K. W. \textit{Appl. Phys. Lett.}
\textbf{2005}, 87, 071103.

(23) Garnett, J. C. M. \textit{Philos. Trans. R. Soc. London, Ser. A
} \textbf{1904}, 203, 385; \textbf{1906}, 205, 237.

(24) Huang, J. P.; Wan, J. T. K.; Lo, C. K.; Yu, K. W. \textit{Phys.
Rev. E} \textbf{2001}, 64, 061505.

(25) Landau L. D.; Lifshitz, E. M.; Pitaevskii, L. P. {\it
Electrodynamics of Continuous Medium,} 2nd ed.; Pergamon Press: New
York, 1984; Chapter 2.

(26) Doyle, W. T.; Jacobs, I. S. \textit{J. Appl. Phys.} {\bf 1992},
71, 3926.

(27) Yu K. W.; Wan, J. T. K. \textit{Comput. Phys. Commun.}
\textbf{2000}, 129, 177.

(28) Huang, J. P.; Karttunen, M.; Yu, K. W.; Dong, L. \textit{Phys.
Rev. E} \textbf{2003}, 67, 021403.

%\end{references}

\newpage

\section*{Figure Captions}

Figure~1.  Cases of nonmagnetic spherical particles. $F$ as a
function of $q$ for different volume fractions, for longitudinal
field cases. Parameters: $\mu_{1}=\mu_0$ and $\mu_{2}=2.9\mu_0$.

\begin{figure}[h]
\includegraphics[width=300pt]{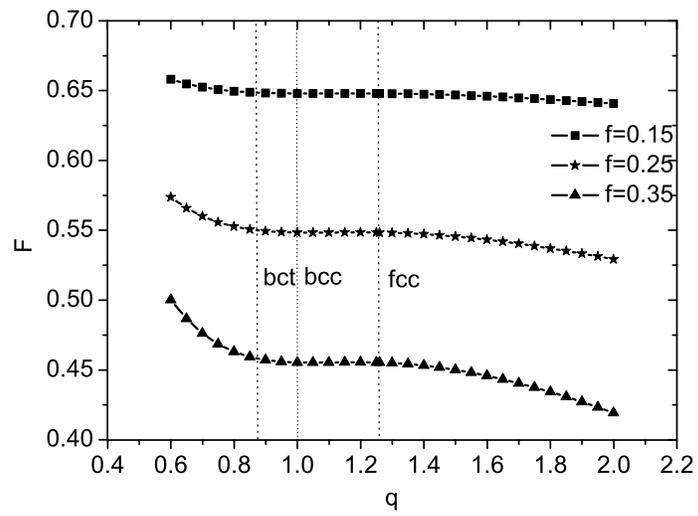} .
\caption{/Gao, Jian, Zhang, and Huang}
\label{fig1}
\end{figure}

\begin{figure}[h]
\includegraphics[width=300pt]{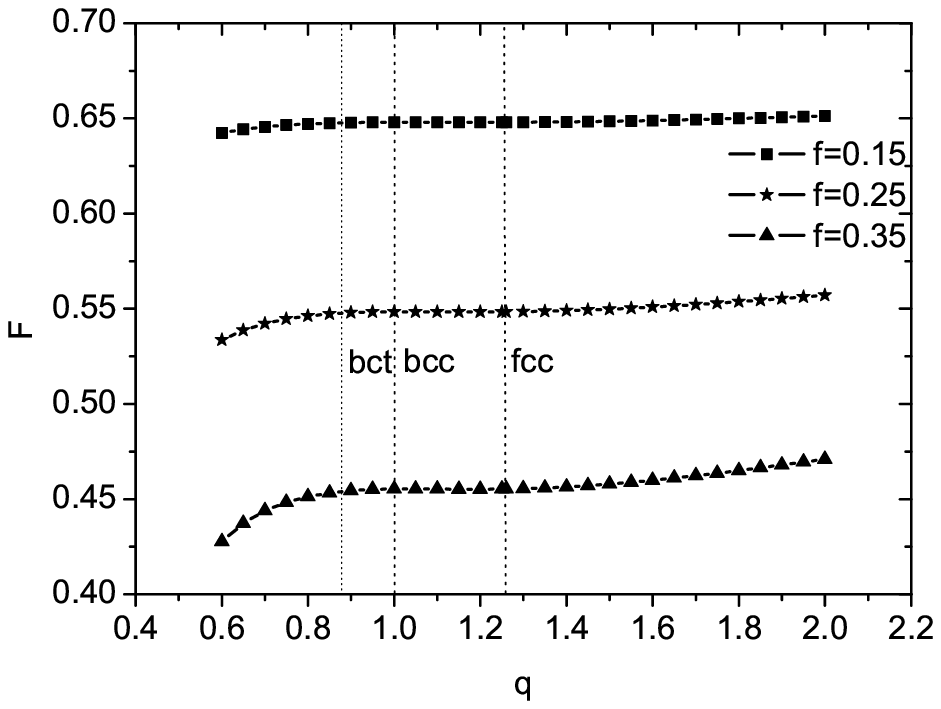} .
\caption{/Gao, Jian, Zhang, and Huang}
\label{fig2}
\end{figure}

Figure~2.  Same as Figure~\ref{fig1}, but for transverse field
cases.

Figure~3.  Cases of nonmagnetic prolate spheroidal particles.
$F_{1}$ as a function of $q$ for different aspect ratios: $k=4,$
$k=6,$ $k=8,$ and $k=10,$ for longitudinal field cases.
Parameters: $\mu_{1}=\mu_0$, $\mu_{2}=2.9\mu_0$, and $f=0.15$.

\begin{figure}[h]
\includegraphics[width=300pt]{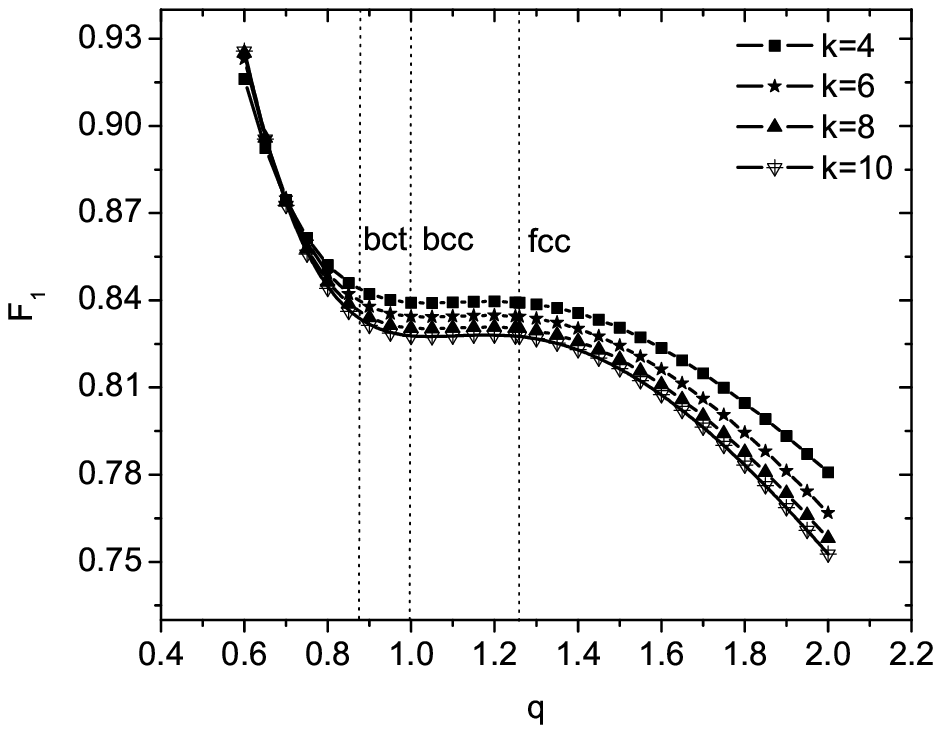} .
\caption{/Gao, Jian, Zhang, and Huang}
\label{fig3}
\end{figure}

\begin{figure}[h]
\includegraphics[width=300pt]{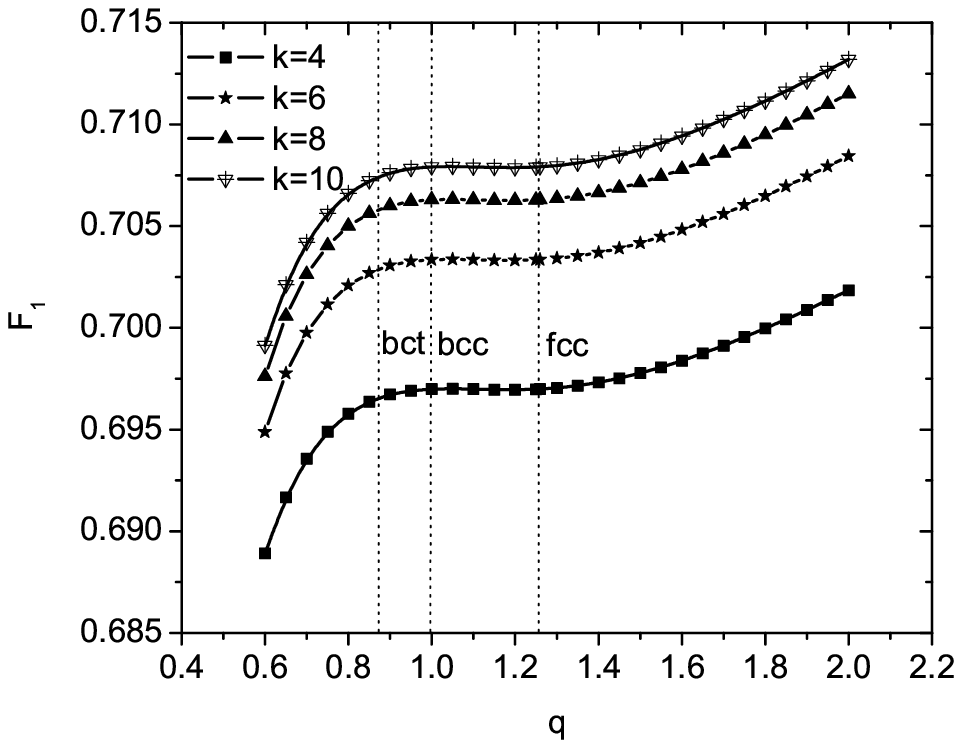} .
\caption{/Gao, Jian, Zhang, and Huang}
\label{fig4}
\end{figure}

Figure~4.  Same as Figure~\ref{fig3}, but for transverse field
cases.

Figure~5.  Cases of nonmagnetic oblate spheroidal particles.
$F_{1}$ as a function of $q$ for different aspect ratios: $k=4,$
$k=6,$ $k=8,$ and $k=10,$ for longitudinal field cases.
Parameters: $\mu_{1}=\mu_0$, $\mu_{2}=2.9\mu_0$, and $f=0.15$.

\begin{figure}[h]
\includegraphics[width=300pt]{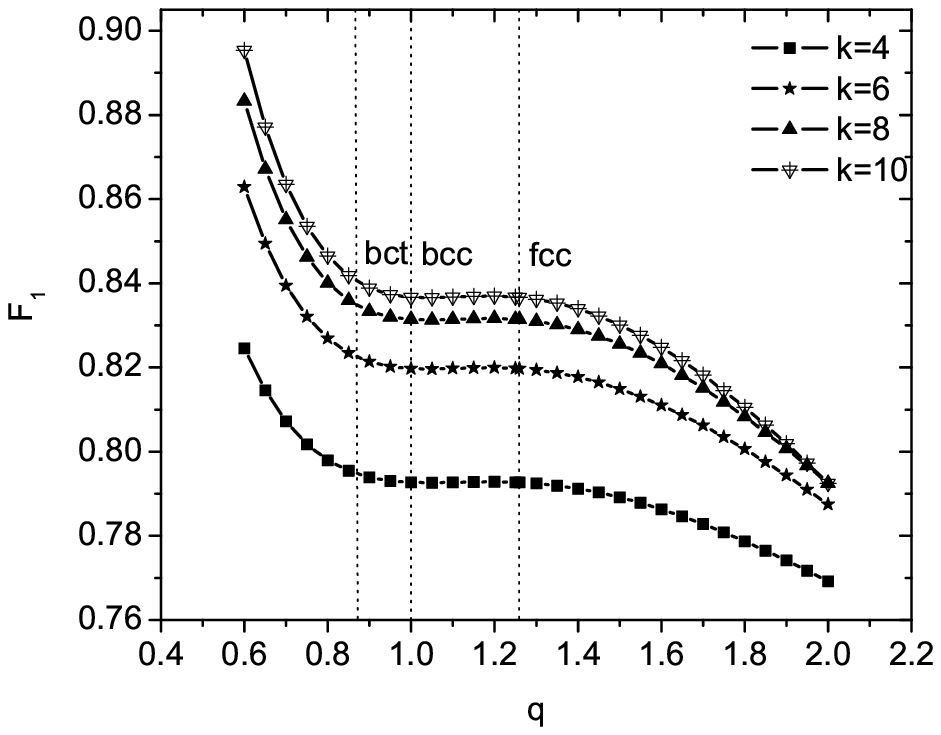} .
\caption{/Gao, Jian, Zhang, and Huang}
\label{fig5}
\end{figure}

\begin{figure}[h]
\includegraphics[width=300pt]{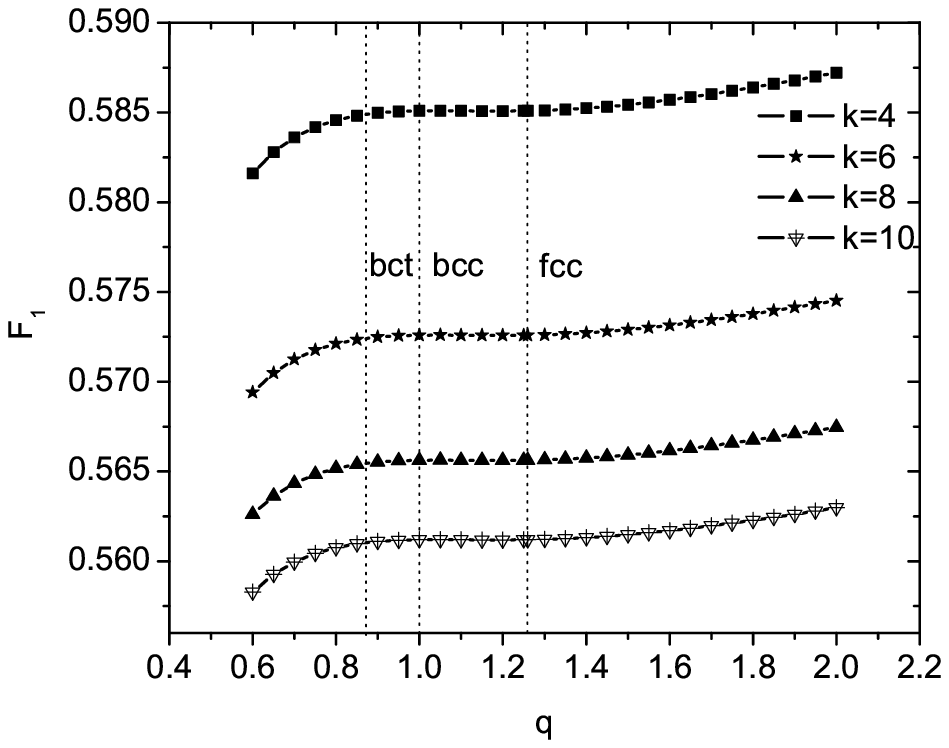} .
\caption{/Gao, Jian, Zhang, and Huang}
\label{fig6}
\end{figure}

\begin{figure}[h]
\includegraphics[width=300pt]{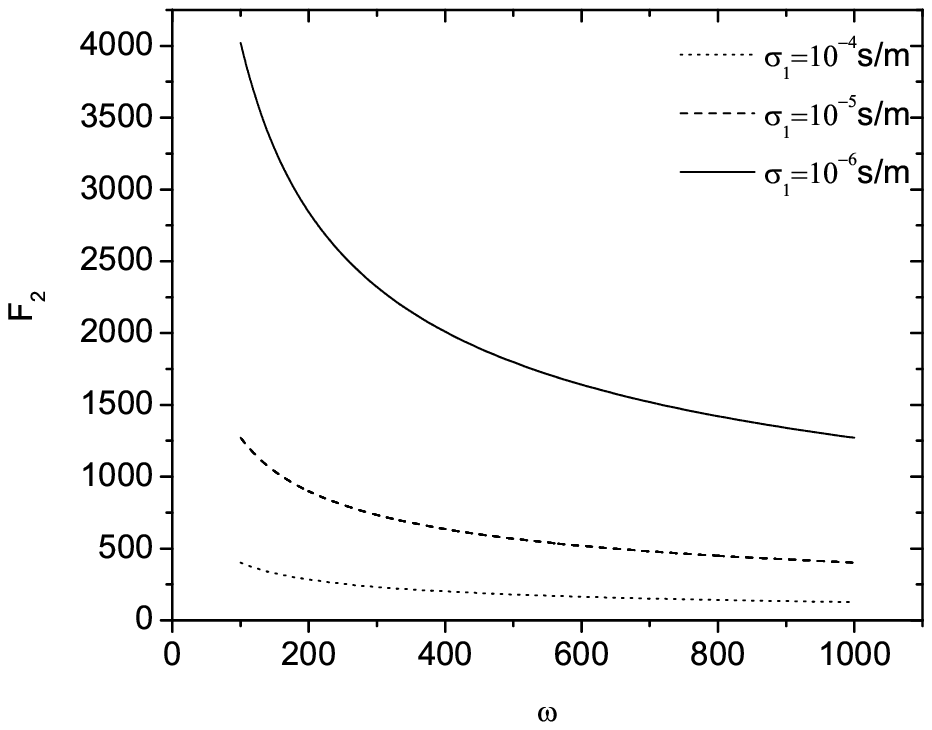} .
\caption{/Gao, Jian, Zhang, and Huang}
\label{fig7}
\end{figure}

Figure~6.  Same as Figure~\ref{fig5}, but for transverse field
cases.

Figure~7.  Cases of nonmagnetic, electrically conductive spherical
particles. $F_{2}$ as a function of $\omega$ for different
conductivities: $\sigma_1=10^{-4}\,$S/m, $10^{-5}\,$S/m, and
$10^{-6}\,$S/m. This figure is for bct lattices, namely,
$q=0.87358$ (which yields $\alpha_L = 1.09298$), and the
longitudinal field cases. Parameters: $\mu_{1}=\mu_{0}$,
$\mu_{2}=2.9\mu_{0}$, $R=2.0\times 10^{-5}\,$m, and $f=0.15.$

\clearpage
\newpage

\clearpage
\newpage

\clearpage
\newpage

\clearpage
\newpage

\clearpage
\newpage

\clearpage
\newpage

\clearpage
\newpage

%%%%%%%%%%%%%%%%%%%%%%%%%%%%%%%%%

\end{document}